\documentclass[10pt,twocolumn,letterpaper]{article}

\usepackage[pagenumbers]{cvpr} 

\usepackage{graphicx}
\usepackage{amsmath}
\usepackage{amssymb}
\usepackage{booktabs}

\usepackage[pagebackref,breaklinks,colorlinks]{hyperref}

\usepackage{wrapfig}

\usepackage[capitalize]{cleveref}
\crefname{section}{Sec.}{Secs.}
\Crefname{section}{Section}{Sections}
\Crefname{table}{Table}{Tables}
\crefname{table}{Tab.}{Tabs.}

\def\cvprPaperID{*****} 
\def\confName{CVPR}
\def\confYear{2022}

\begin{document}

\title{BundleFit: Display and See-Through Models for Augmented Reality Head-Mounted Displays}

\author{Yufeng Zhu\\
{\tt\small mike323zyf@gmail.com}
}
\maketitle

\begin{abstract}
The head-mounted display is a vital component of augmented reality, incorporating optics with complex display and see-through optical behavior. Computationally modeling these optical behaviors requires meeting three key criteria: accuracy, efficiency, and accessibility. In recent years, various approaches have been proposed to model display and see-through optics, which can broadly be classified into black-box and white-box models. However, both categories face significant limitations that hinder their adoption in commercial applications. To overcome these challenges, we leveraged prior knowledge of ray bundle properties outside the optical hardware and proposed a novel bundle-fit-based model. In this approach, the ray paths within the optics are treated as a black box, while a lightweight optimization problem is solved to fit the ray bundle outside the optics. This method effectively addresses the accuracy issues of black-box models and the accessibility challenges of white-box models. Although our model involves runtime optimization, this is typically not a concern, as it can use the solution from a previous query to initialize the optimization for the current query. We evaluated the performance of our proposed method through both simulations and experiments on real hardware, demonstrating its effectiveness.
\end{abstract}

\section{Introduction}
\label{sec:intro}

In augmented reality (AR), computer-generated visual content is superimposed on the user's view of the real world through devices such as smartphones, tablets, and specialized glasses. An AR head-mounted display (HMD), typically consisting of a see-through display placed in front of the user's eyes, is a device worn on the head that allows the user to see both the real world and virtual content overlaid on it. AR HMDs can be used for a wide range of applications, including entertainment, education, and industrial training. They have the potential to revolutionize the way we interact with the world and each other by providing a more immersive and interactive experience. Some examples of AR HMDs include the Microsoft HoloLens and the Magic Leap One \cite{ASO:ART1997,RTI:KBB2018}.

To achieve a seamless and immersive AR experience, the virtual elements must be correctly registered with the physical objects. This is where world-locked rendering (WLR) plays a crucial role, as it is a technique used in AR systems to generate visual content that appears as if it is part of the real world, rather than being displayed on top of it. WLR systems rely on a combination of sensors and computer vision techniques to ensure that the virtual content is correctly aligned even if the users move or change their perspective. Despite the advances made in WLR, there are still a number of challenges that need to be addressed, which includes improving the accuracy and stability of various sub-modules, such as inside-out tracking (IOT), eye tracking, display and see-through modeling, etc \cite{TIA:ILS2021}. In this work, we focus on addressing the topics related to the display and see-through modeling of AR HMDs with complex optical designs.

The optical design of a see-through display can present various challenges for the WLR system, including highly nonlinear geometric distortion, non-uniformity, and other issues. Additionally, changes in the user's pupil position can exacerbate these challenges due to the phenomenon known as "pupil swim" \cite{VOF:GGW2018}. To address the problem of viewpoint-dependent geometric distortion, researchers have investigated various approaches to display and see-through modeling \cite{ASO:GIM2018}. However, they have struggled with either accuracy issues or a lack of flexibility. In this paper, we propose new display and see-through models that are designed to overcome these difficulties. Briefly, we examine the fiber bundle structure \cite{FB:HD1996} within the context of display and see-through problems and present black box fitting models that can be generalized to different optical designs of see-through displays and exhibit improved structure preservation properties, leading to enhanced accuracy. We will begin by discussing our proposed models, and then outline the process of calibration and WLR systems integration. Finally, we assess the performance of the models through both simulated experiments and real-world evaluations using AR HMDs. As we will demonstrate, our models are not dependent on the specific optical design and are able to achieve high levels of accuracy in modeling performance.

\section{Related Work}
\label{sec:related_work}

Our research falls under the category of AR HMD display and see-through calibrations, a highly active area of study within the domains of computer vision and graphics \cite{ASO:GIM2018}. Historically, this task has been challenging due to the complexity of the optics involved \cite{ISA:AB1994,FCF:FSP1999,AMF:MT1999,OST:GSW2000,GCO:KBB2012}. Traditional methods, such as the Single Point Active Alignment Method (SPAAM) and its variations \cite{SPA:TN2000,PSF:GTN2002,CUS:GTM2010} were used to estimate the projection from the real world to the display by requiring users to manually align highlighted pixels with pre-determined 3D points, which can be time-consuming and susceptible to user errors. To minimize human involvement in the calibration process, Owen et al. \cite{DRC:OZA2004} proposed a two-phase method that greatly reduces manual alignment effort. More recently, fully automated approaches \cite{SCO:GFG2008,IFC:IK2014} have been developed to eliminate the need for manual calibration altogether, making the process more efficient and less prone to errors caused by user variability. However, these methods have limitations in terms of accuracy \cite{PAS:IK2014}, as they often rely on simplified models for the display and see-through optics. Additionally, the geometric distortion caused by the optics can vary significantly depending on the user's pupil position, known as pupil swim \cite{VOF:GGW2018,TCO:F1962,ODO:CKA2002}, which also needs to be considered in the calibration process.

To improve the accuracy of WLR systems, different methods have been examined. Recognizing that each display pixel is perceived as a point light source from the viewer's perspective, researchers have employed triangulation techniques to determine its 3D position for each pixel in order to model the display optics \cite{NPC:KSH2016,HAP:KSH2017}. However, this assumption is not sufficient to fully capture complex optical designs, as the rays corresponding to each pixel may not converge at a single point. As an alternative, using a light field representation is more appropriate for precise modeling \cite{LFC:IK2015,SDA:IK2015}. Another direction is to trace the rays through display and see-through optics for each pixel and estimate unknown parameters by minimizing reprojection errors \cite{RCF:GTS2020,VOF:GGW2018}. To reduce the computational cost of ray tracing, researchers have applied nonlinear dimension reduction techniques to pre-traced light rays from simulations, enabling real-time image distortion correction \cite{PRF:GMS2022}. These methods, however, are sensitive to the accuracy of manufacture and assembly process, as they depend on a certain Computer-Aided Design (CAD) model. Additionally, they are inflexible as the ray tracing software implementation is specific to certain optical systems. With the growing trend of neural network development, this technique has also been adopted in the modeling of display optics \cite{OCM:SHY2019,NDF:HSI2022}.

\section{Display And See-Through Models}
\label{sec:main}

\begin{figure}[t]
  \centering
   \includegraphics[width=\linewidth]{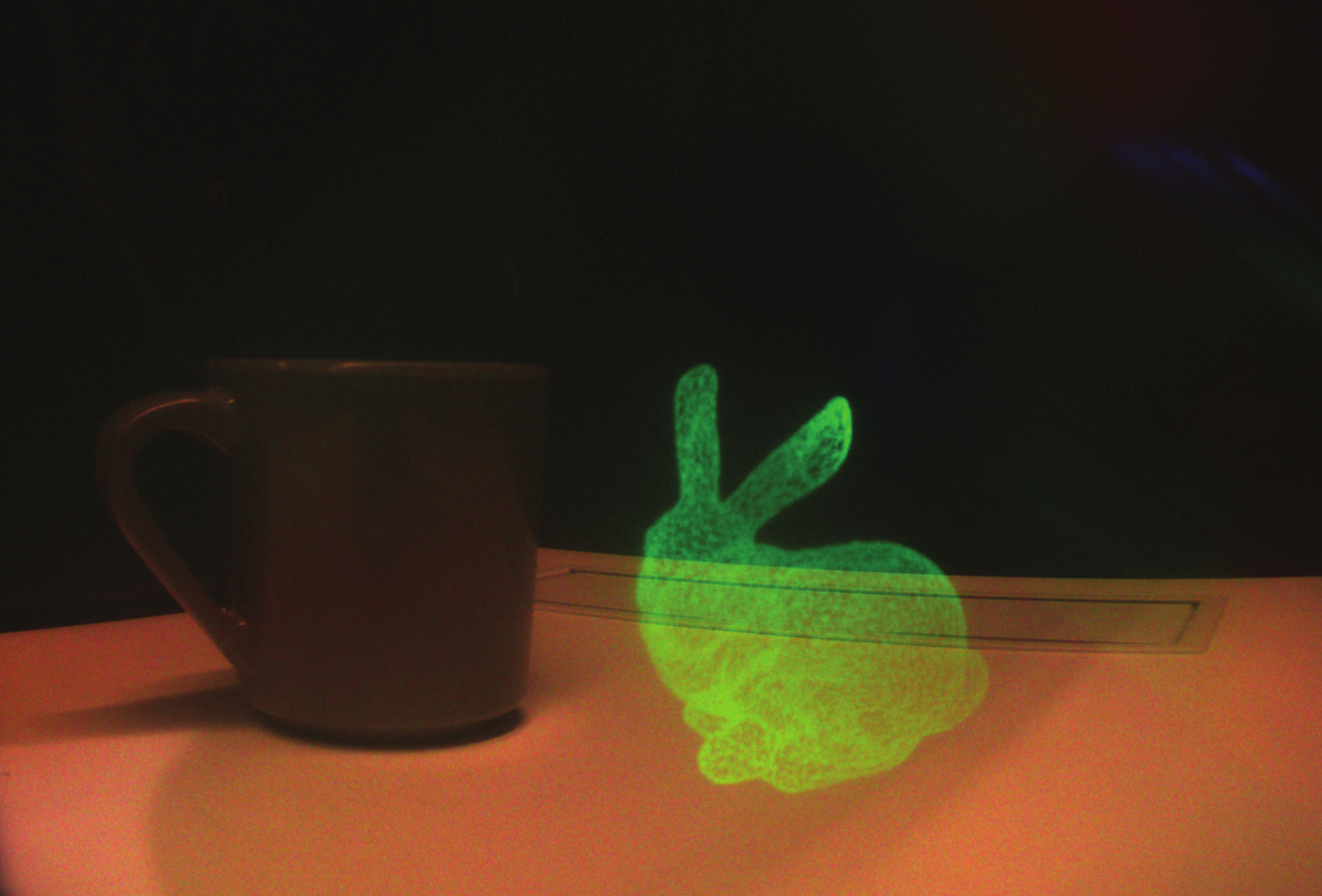}
   \caption{Real example figure for World Locked Rendering.}
   \label{fig:wlr_example}
\end{figure}

Display and see-through models are essential components for world-locked AR, which is to position virtual objects with reference to real objects as shown in \cref{fig:wlr_example}. The display model is designed to correct for the complex geometric distortions that are often introduced by the display optics, such as a curved combiner. The see-through model is needed to align the display and IOT subsystem within a common coordinate frame. To align virtual and real objects accurately, the AR system must have a precise mapping from the physical world to the display panel. This is achieved by performing a calibration process for the IOT, eye tracking, display, and see-through models prior to using world-locked applications on AR devices. In this paper, we will focus on the discussion of the display and see-through models, which are the main technical contributions of our work. We will also briefly mention the IOT and eye tracking modules to show how our proposed models can be integrated into a world-locked AR system. We direct readers interested in these topics to further literature \cite{PR:TS2002,MVG:HZ2003,ETA:HNA2011} for a more comprehensive review.

\subsection{Background}
\label{sec:background}

\begin{figure}[b]
  \centering
   \includegraphics[width=\linewidth]{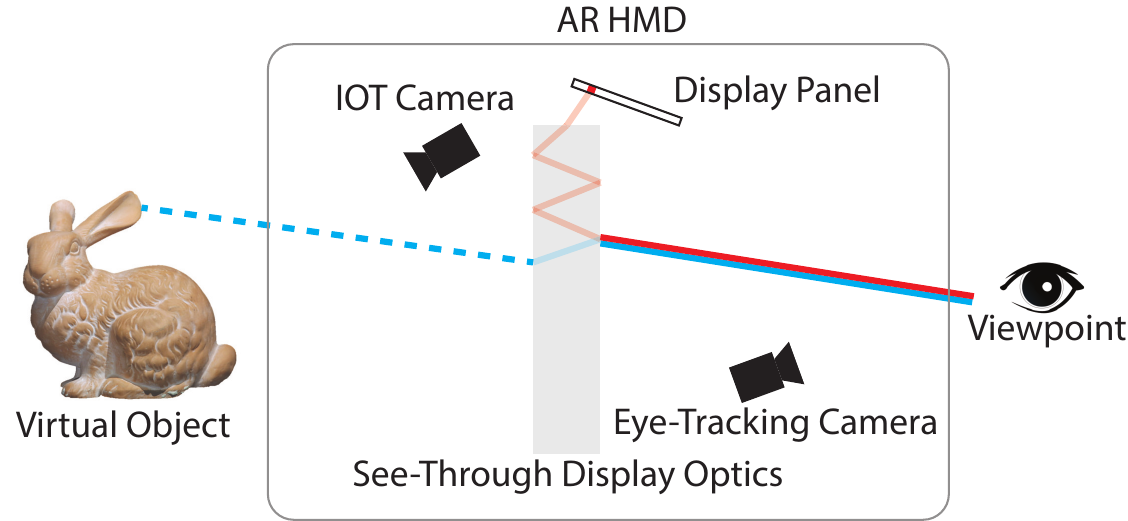}
   \caption{Explaination figure for World Locked Rendering.}
   \label{fig:wlr_explaination}
\end{figure}

We begin with a brief introduction to the WLR system and emphasize the roles of the display and see-through models. For the sake of simplicity, we make the following assumptions: (1) camera's depth of field is infinite; (2) we are limiting our consideration to geometric optics; (3) undesired artifacts caused by the optics, such as ghosting, non-uniformity, and others, are not taken into account. By using an HMD with rigidly mounted IOT cameras and IMUs, we can estimate the HMD's pose through front-end visual inertial odometry (VIO) and the geometry of the surrounding physical objects through back-end simultaneously localization and mapping (SLAM), provided that the IOT system is calibrated. The IOT problem has been extensively researched in previous works \cite{AMS:MRO2007,HPC:LM2013}. An HMD equipped with an IOT system will have both its pose and the geometry of the real world represented in its own coordinate frame (i.e. device frame). Without loss of generality, we will consider the monocular case only. As an illustration, we have selected a sculpture of a bunny, shown in \cref{fig:wlr_explaination}. Our goal is to demonstrate how to render a virtual bunny on the display in such a manner that it appears aligned with the real sculpture from the perspective of a viewpoint provided by the eye tracking system. It is assumed that the eye tracking cameras are calibrated and connected to the IOT system, allowing both the geometry of the sculpture and the viewpoint to be represented in the device frame.

To determine the values of each pixel on the display panel, researchers have suggested using a white box model \cite{RCF:GTS2020,VOF:GGW2018}, which involves explicitly tracing rays throughout the optical system. As depicted in \cref{fig:wlr_explaination}, the path of the corresponding ray (\textcolor{red}{red}) is traced from each pixel through the specified viewpoint, which reveals how the pixel is perceived from that perspective. Next, a ray (\textcolor{blue}{blue}) with the same viewing angle is traced through the see-through optics, emanating from the viewpoint and projecting into the real world. Eventually, the intersection of this ray path (\textcolor{blue}{blue}) with the virtual objects to be displayed (in this case, the virtual geometry of the sculpture computed by the SLAM) determines the value of the corresponding pixel.

The white box model has the advantage of being physically interpretable, but suffers from the following drawbacks: (1) it relies on the accuracy of the hardware manufacturing and assembly process in order to achieve accurate performance; (2) it lacks flexibility and requires changes to the software implementation when the hardware design is altered. Alternatively, a black box model can avoid these issues by modeling the relationship between the display pixels and the rays (\textcolor{blue}{dashed blue}) projected into the real world, which is ultimately what matters, directly as a function (see \cref{fig:wlr_explaination}) and omit all the intermediate details of ray tracing. However, this approach is not generally effective as it disregards the role of the viewer's perspective. In general, the pixel-ray mapping is viewpoint dependent because the correspondence it describes varies significantly according to the viewer's positions due to the effect of pupil swim \cite{VOF:GGW2018}.

To address this issue, we utilize two black box models: the display model and the see-through model, eliminating the need for intricate ray tracing process within display and see-through optics (\textcolor{red}{light red} and \textcolor{blue}{light blue}). The display model (in \cref{sec:display}) explains the relationship between the display pixels and the rays (\textcolor{red}{solid red}) that are perceived from the viewer's perspective, while the see-through model (in \cref{sec:seethru}) describes the correspondence between rays (\textcolor{blue}{solid blue}) starting from the viewpoint and rays (\textcolor{blue}{dashed blue}) that project into the real world. In order to provide a better understanding of the motivation and context for our new approach, we begin by discussing the rationale behind it in \cref{sec:fiber_bundle_fitting}.


\subsection{Fiber Bundle Fitting}
\label{sec:fiber_bundle_fitting}

Model fitting is a well-researched technique that has been widely applied to solve a diverse range of problems and challenges in various fields. It involves identifying the most suitable model to describe the relationships between a set of variables, estimating the model's parameters using available data, and evaluating the model's ability to represent the relationships and make predictions \cite{MLA:MKP2012}. In this work, we are applying this technique to address display and see-through problems, attempting to find models that can accurately describe the correspondence relationships involving pixels and rays. 

One of the primary contributions of this work is the finding that, when the data space exhibits a fiber bundle structure and the projection operation is feasible, it is more effective to fit the subset of data on the base space rather than the entire data space in order to better preserve the fiber bundle structure. For example, let's assume that we have a set of fitting data, $(p_i, q_i), p_i \in P, q_i \in Q$, and want to fit the correspondence model, $f: P \mapsto Q$. If the data space $P$ has a fiber bundle structure $\pi: P \mapsto B$, where $P$ is the total space and $B$ is the base space, we can fit intermediate model $\tilde{f}: B \mapsto Q$ instead. Then the desired models can be constructed as $f = \tilde{f}\circ\pi$ by combining with the projection operation $\pi$. Inspired by this idea, we will exploit the fiber bundle structure property in the development of the display and see-through models.


\subsection{Display Model}
\label{sec:display}

\begin{figure}[t]
  \centering
   \includegraphics[width=\linewidth]{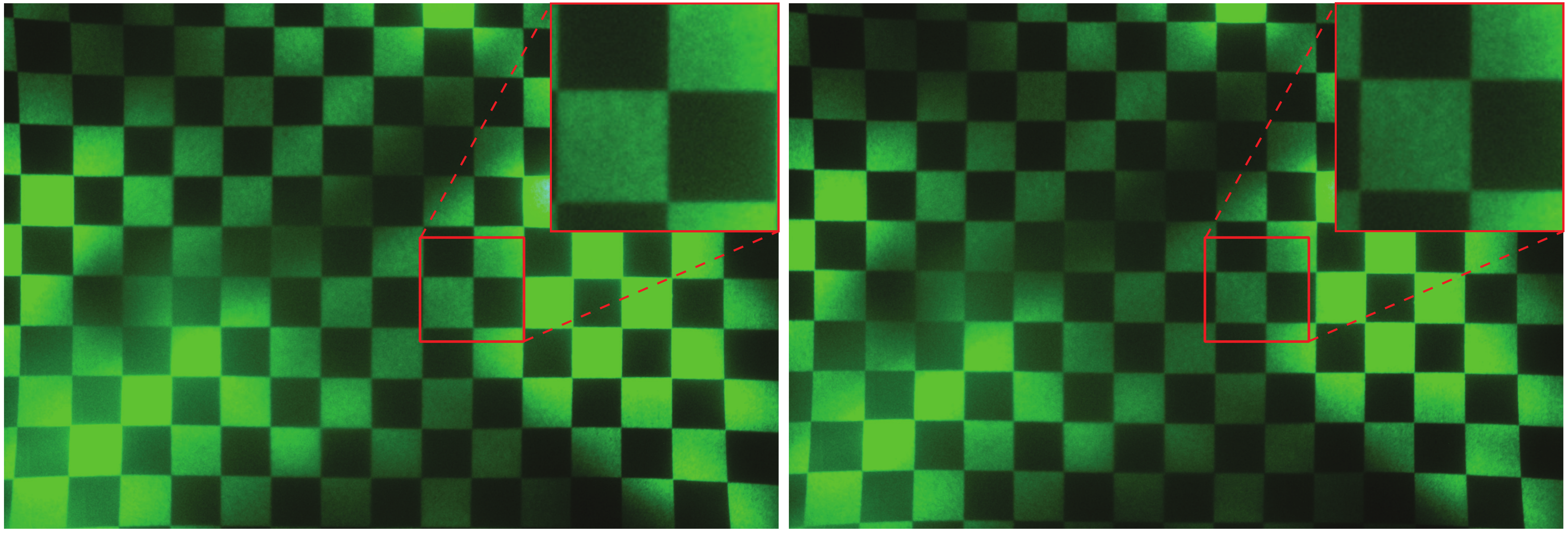}
   \caption{Pupil swim for display model.}
   \label{fig:display_pupil_swim}
\end{figure}

The display model illustrates the connection between the display pixels and the rays perceived from the viewer's perspective. This correspondence may vary as the viewpoint shifts due to the effect of pupil swim as shown in \cref{fig:display_pupil_swim}. Thus the forward and backward display problems can be defined as
\begin{equation}
\begin{split}
  f_{display}:(\textbf{p}_{pixel}, \textbf{p}_{view})\mapsto \textbf{v}_{view},\\
  \textbf{p}_{pixel}\in\mathbb{R}^2, \textbf{p}_{view}\in\mathbb{R}^3, \textbf{v}_{view}\in\mathbb{S}^2,
  \label{eq:display_forward}
\end{split}
\end{equation}
and
\begin{equation}
\begin{split}
  f^{-1}_{display}:(\textbf{p}_{view}, \textbf{v}_{view})\mapsto \textbf{p}_{pixel},
  \label{eq:display_backward}
\end{split}
\end{equation}
respectively. In \cref{eq:display_forward} and \cref{eq:display_backward}, $\textbf{p}_{pixel}$ denotes a display pixel's position, while $\textbf{p}_{view}$ is the viewpoint position and $\textbf{v}_{view}$ represents the direction of the corresponding perceived ray.

\begin{figure}[t]
  \centering
   \includegraphics[width=\linewidth]{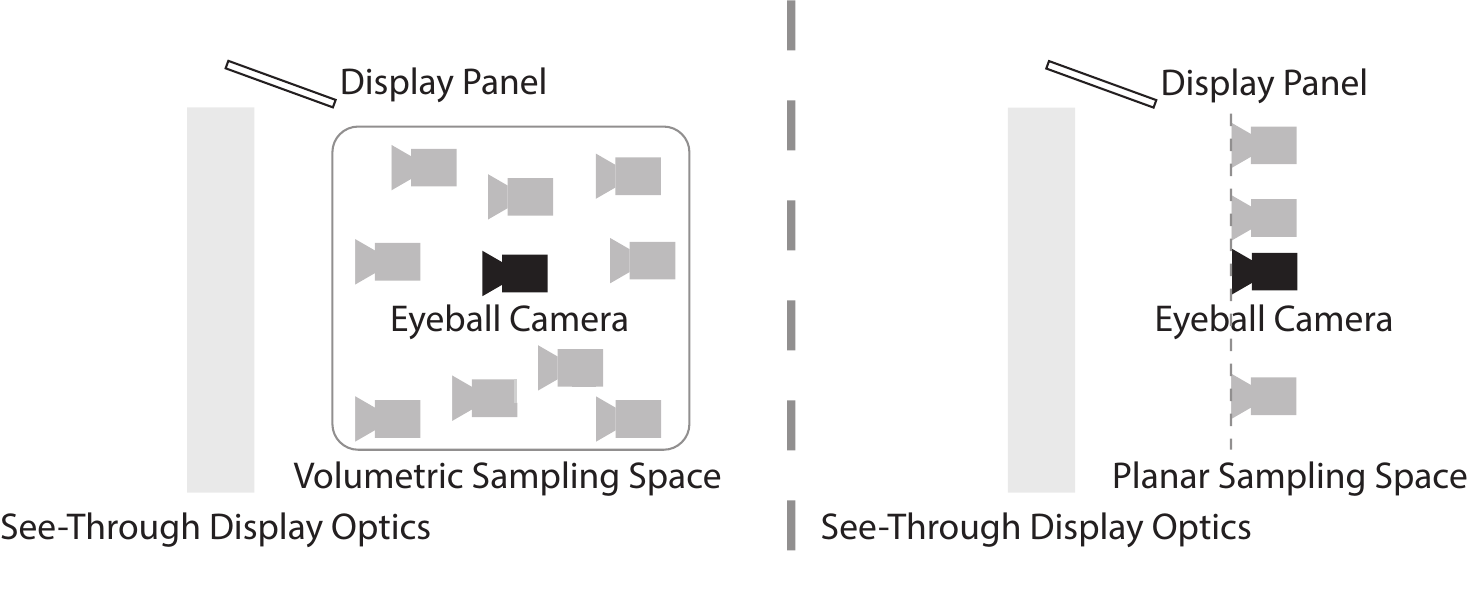}
   \caption{Volume sampling approach vs plane sampling.}
   \label{fig:sampling_comparison}
\end{figure}

\begin{figure}[b]
     \centering
     \begin{subfigure}[b]{0.23\textwidth}
         \centering
         \includegraphics[width=\textwidth]{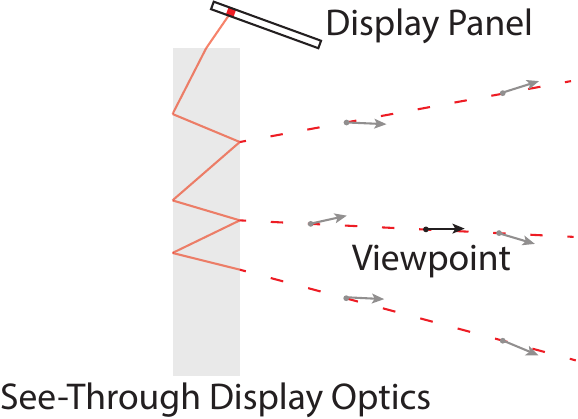}
         \caption{Volume sampling approach failure for display.}
         \label{fig:volumetric_failure_1}
     \end{subfigure}
     \hfill
     \begin{subfigure}[b]{0.23\textwidth}
         \centering
         \includegraphics[width=\textwidth]{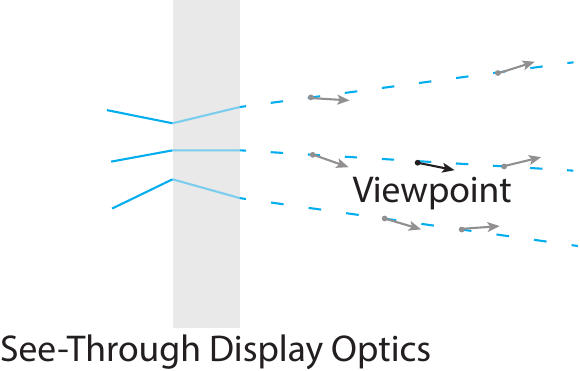}
         \caption{Volume sampling approach failure for seethru.}
         \label{fig:volumetric_failure_2}
     \end{subfigure}
        \caption{Volume sampling approach failure.}
        \label{fig:volumetric_failure}
\end{figure}

\paragraph{Volumetric Viewpoint Sampling}
It is straightforward to model such correspondence by following the problem definitions. To collect data for fitting the model, a calibrated eyeball camera is used to represent the viewer's perspective. This process involves sampling pixels on the 2D display panel and taking volumetric samples of viewpoints by placing the eyeball camera at various 3D locations to capture the corresponding rays from different perspectives (see \cref{fig:sampling_comparison}). Parameters of the chosen model can then be estimated using the collected data. However, this volumetric viewpoint sampling approach is resource-intensive and prone to inaccurate modeling. Specifically, if the fitting step is not carefully constrained, the fitted model may incorrectly predict that two viewpoints with the same viewing angle for a pixel will have different perspectives of that pixel (see \cref{fig:volumetric_failure}).

\begin{figure}[t]
     \centering
     \begin{subfigure}[b]{0.23\textwidth}
         \centering
         \includegraphics[width=\textwidth]{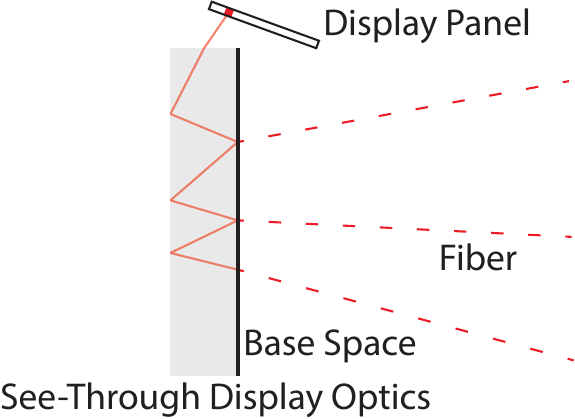}
         \caption{Display fiber bundle structure.}
         \label{fig:display_fiber_bundle_1}
     \end{subfigure}
     \hfill
     \begin{subfigure}[b]{0.23\textwidth}
         \centering
         \includegraphics[width=\textwidth]{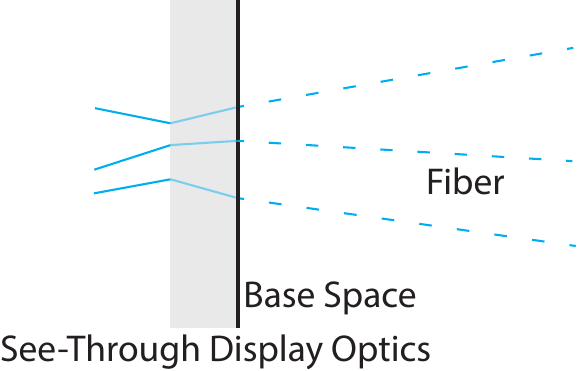}
         \caption{Seethru fiber bundle structure.}
         \label{fig:display_fiber_bundle_2}
     \end{subfigure}
        \caption{Fiber bundle structure.}
        \label{fig:display_fiber_bundle}
\end{figure}

\paragraph{Planar Viewpoint Sampling}
Upon closer examination of the correspondence problems in \cref{eq:display_forward} and \cref{eq:display_backward}, we observe that the data space $(\textbf{p}_{view}, \textbf{v}_{view})$ of each pixel, which consists of its corresponding rays emanating from the display optics, has a fiber bundle structure. As shown in \cref{fig:display_fiber_bundle}, the base space for each pixel's fiber bundle is the surface of the display optics and the fibers are its corresponding rays. Motivated by the discussion in \cref{sec:fiber_bundle_fitting}, we introduce two intermediate model problems in a similar way,
\begin{equation}
\begin{split}
  \tilde{f}_{display}:(\textbf{p}_{pixel}, \tilde{\textbf{p}}_{view})\mapsto \textbf{v}_{view},\\
  \textbf{p}_{pixel}, \tilde{\textbf{p}}_{view}\in\mathbb{R}^2, \textbf{v}_{view}\in\mathbb{S}^2,
  \label{eq:display_intermediate_forward}
\end{split}
\end{equation}
and
\begin{equation}
\begin{split}
  \tilde{f}^{-1}_{display}:(\tilde{\textbf{p}}_{view}, \textbf{v}_{view})\mapsto \textbf{p}_{pixel}.
  \label{eq:display_intermediate_backward}
\end{split}
\end{equation}
Our proposed method involves sampling the viewpoint position $\tilde{\textbf{p}}_{view}$ on a two-dimensional plane, $\mathbb{P}_{display}$, as depicted in \cref{fig:sampling_comparison}, rather than in three dimensions. We then use the projection operations to recover the desired forward and backward display models.

The projection operation $\pi_{display}^{\vee}$ for the backward model is simple and involves intersecting the 2D plane $\mathbb{P}_{display}$ with the ray emanating from the 3D viewpoint $\textbf{p}_{view}$ and traveling in the opposite direction of $\textbf{v}_{view}$ as in \cref{eq:display_projection}.
\begin{equation}
\begin{split}
  \pi_{display}^{\vee}:(\textbf{p}_{view}, \textbf{v}_{view})\mapsto \tilde{\textbf{p}}_{view},\\
  \tilde{\textbf{p}}_{view} = \textbf{p}_{view} - t^* \cdot \textbf{v}_{view},
  \label{eq:display_projection}
\end{split}
\end{equation}
where
\begin{equation}
\begin{split}
  t^* = \underset{t \geq 0}{\mathrm{argmin}}\, dist(\textbf{p}_{view} - t \cdot \textbf{v}_{view}, \mathbb{P}_{display}),
  \label{eq:display_projection_expansion}
\end{split}
\end{equation}
and $dist(\cdot, \cdot)$ is the distance function. Therefore, the backward display model can be constructed as
\begin{equation}
\begin{split}
  f^{-1}_{display} = \tilde{f}^{-1}_{display}(\pi_{display}^{\vee}(\textbf{p}_{view}, \textbf{v}_{view}), \textbf{v}_{view}).
  \label{eq:display_backward_constructed}
\end{split}
\end{equation}

While the projection operation for the backward model is relatively obvious, the forward model is more involved. In the forward case, the goal is to predict how a pixel $\textbf{p}_{pixel}$ is perceived from a given viewpoint $\textbf{p}_{view}$, but the intermediate forward model $\tilde{f}_{display}$ that we fit requires the viewpoint position to be located on the base plane $\mathbb{P}_{display}$. We formulate this projection operation, $\pi_{display}^{\wedge}$, as an optimization problem,
\begin{equation}
\begin{split}
 \pi_{display}^{\wedge}&:(\textbf{p}_{pixel}, \textbf{p}_{view})\mapsto \tilde{\textbf{p}}_{view},\\
 \tilde{\textbf{p}}_{view}^*, t^* &= \underset{\tilde{\textbf{p}}_{view}, t \geq 0}{\mathrm{argmin}}\, dist(g(\tilde{\textbf{p}}_{view}, \textbf{p}_{pixel}, t), \textbf{p}_{view}),\\
 &s.t.\,\,\,\, \tilde{\textbf{p}}_{view} \in \mathbb{P}_{display},\\
 g(\tilde{\textbf{p}}_{view}&, \textbf{p}_{pixel}, t) = \tilde{\textbf{p}}_{view} + t \cdot \tilde{f}_{display}(\textbf{p}_{pixel}, \tilde{\textbf{p}}_{view}),
  \label{eq:display_lifting_optimization}
\end{split}
\end{equation}
whose global optimal candidate $\tilde{\textbf{p}}_{view}^*$ is the projection of $\textbf{p}_{view}$. Or in other words, $\tilde{\textbf{p}}_{view}^*$ perceives $\textbf{p}_{pixel}$ in the same way as $\textbf{p}_{view}$ perceives it. Then it is straightforward to construct the forward display model as follows:
\begin{equation}
\begin{split}
  f_{display} = \tilde{f}_{display}(\textbf{p}_{pixel}, \pi_{display}^{\wedge}(\textbf{p}_{pixel}, \textbf{p}_{view})).
  \label{eq:display_forward_constructed}
\end{split}
\end{equation}
It is evident that our new approach is more resource-efficient compared to the volumetric viewpoint sampling approach. More importantly, it avoids the fitting artifacts, as illustrated in \cref{fig:volumetric_failure}, by leveraging the fiber bundle structure present in the display problems.

\subsection{See-Through Model}
\label{sec:seethru}

\begin{figure}[b]
  \centering
   \includegraphics[width=\linewidth]{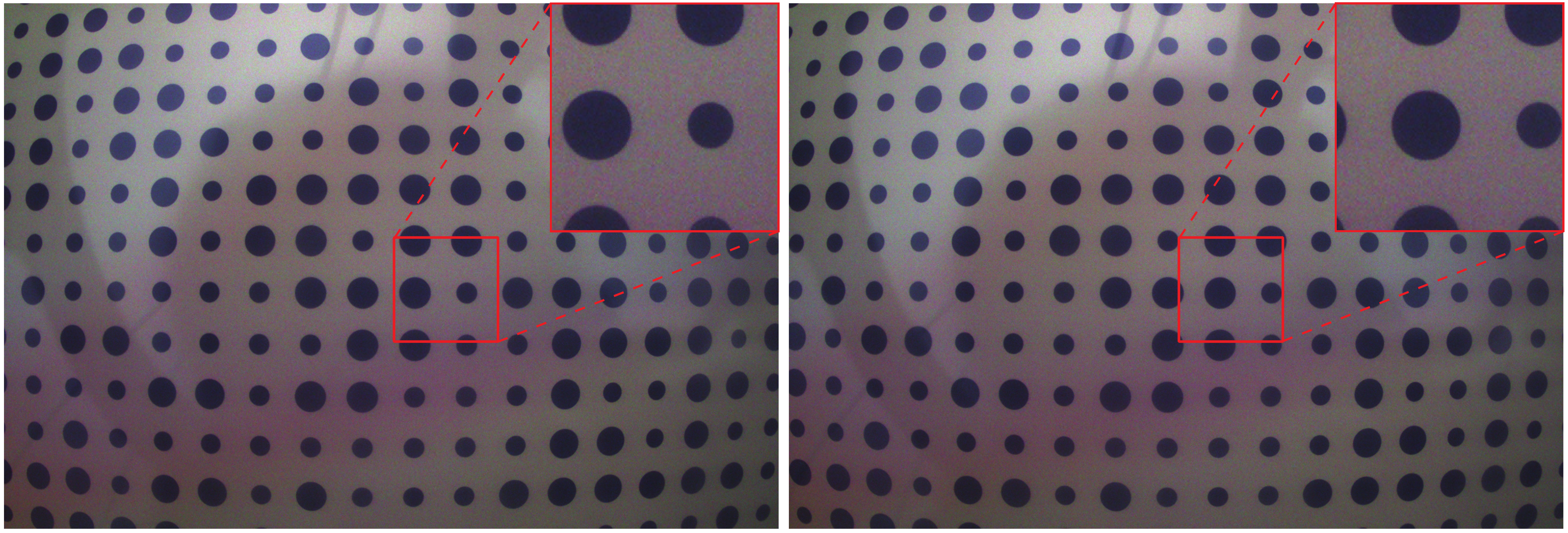}
   \caption{Pupil swim for see-through model.}
   \label{fig:seethru_pupil_swim}
\end{figure}

The see-through model describes the relationship between rays that originate from the viewpoint and rays that project into the real world. Similar to the display model, it also depends on the viewpoint, as shown in \cref{fig:seethru_pupil_swim}. The forward and backward see-through problems can be defined as
\begin{equation}
\begin{split}
  f_{seethru}:(\textbf{p}_{view}, \textbf{v}_{view})\mapsto (\textbf{p}_{real}^+, \textbf{p}_{real}^-),\\
  \textbf{p}_{real}^+, \textbf{p}_{real}^-\in\mathbb{R}^2, \textbf{p}_{view}\in\mathbb{R}^3, \textbf{v}_{view}\in\mathbb{S}^2,
  \label{eq:seethru_forward}
\end{split}
\end{equation}
and
\begin{equation}
\begin{split}
  f^{-1}_{seethru}:&(\textbf{p}_{view}, \textbf{p}_{real})\mapsto \textbf{v}_{view},\\
  &\textbf{p}_{real}\in\mathbb{R}^3,
  \label{eq:seethru_backward}
\end{split}
\end{equation}
respectively. In \cref{eq:seethru_forward} and \cref{eq:seethru_backward}, $\textbf{p}_{view}$ denotes the viewpoint position and $\textbf{v}_{view}$ represents the ray direction starting from the viewpoint as in \cref{sec:display}. Additionally, $\textbf{p}_{real}$ is a 3D location in the real world, while $\textbf{p}_{real}^+$ and $\textbf{p}_{real}^-$ are the two-plane parameterization of the real world light field \cite{TLU:GGS1996,LFR:MP1996}, which are two points located on two real world planes, $\mathbb{P}_{real}^+$ and $\mathbb{P}_{real}^-$. The forward model unravels how the see-through optics alters the path of a ray when it is projected from a viewpoint into the real world. And the backward model details how a real world point will be perceived from a viewer's perspective through the see-through optics. As discussed in \cref{sec:display}, we should avoid volumetric viewpoint sampling approach and take advantage of the fiber bundle structure in the see-through problems. Accordingly, we also adopt a planar viewpoint sampling strategy and propose an intermediate model problem,
\begin{equation}
\begin{split}
  \tilde{f}_{seethru}:(\tilde{\textbf{p}}_{view}, \textbf{v}_{view})\mapsto (\textbf{p}_{real}^+, \textbf{p}_{real}^-),\\
  \tilde{\textbf{p}}_{view}, \textbf{p}_{real}^+, \textbf{p}_{real}^-\in\mathbb{R}^2, \textbf{v}_{view}\in\mathbb{S}^2.
  \label{eq:seethru_forward_intermediate}
\end{split}
\end{equation}
The viewpoint $\tilde{\textbf{p}}_{view}$ is sampled on a 2D plane, $\mathbb{P}_{seethru}$ and we will define the projection operations to restore the forward and backward see-through models.

This time, the forward see-through model shares a similar projection operation with the backward display model (see \cref{eq:display_projection} and \cref{eq:display_projection_expansion}), which is defined as
\begin{equation}
\begin{split}
  \pi_{seethru}^{\wedge}:(\textbf{p}_{view}, \textbf{v}_{view})\mapsto \tilde{\textbf{p}}_{view},\\
  \tilde{\textbf{p}}_{view} = \textbf{p}_{view} - t^* \cdot \textbf{v}_{view},
  \label{eq:seethru_projection}
\end{split}
\end{equation}
where
\begin{equation}
\begin{split}
  t^* = \underset{t \geq 0}{\mathrm{argmin}}\, dist(\textbf{p}_{view} - t \cdot \textbf{v}_{view}, \mathbb{P}_{seethru}).
  \label{eq:seethru_projection_expansion}
\end{split}
\end{equation}
And the forward model can be easily constructed as
\begin{equation}
\begin{split}
  f_{seethru} = \tilde{f}_{seethru}(\pi_{seethru}^{\wedge}(\textbf{p}_{view}, \textbf{v}_{view}), \textbf{v}_{view}).
  \label{eq:seethru_forward_constructed}
\end{split}
\end{equation}
On the other hand, the projection operation of the backward model requires a little bit more effort to build. In this case, we want to understand how a given viewpoint $\textbf{p}_{view}$ perceives a 3D real world point $\textbf{p}_{real}$. Since we've already fitted an intermediate model $\tilde{f}_{seethru}$ in \cref{eq:seethru_forward_intermediate}, which explains how rays from the real world become rays that reach the viewer after passing through the see-through optics, the only remaining task is to determine which ray emanating from $\textbf{p}_{real}$ will ultimately reach $\textbf{p}_{view}$. We formulate this problem as an optimization process,
\begin{equation}
\begin{split}
 &\pi_{seethru}^{\vee}:(\textbf{p}_{view}, \textbf{p}_{real})\mapsto \tilde{\textbf{p}}_{view},\\
 \tilde{\textbf{p}}_{view}^*, t^* &= \underset{\tilde{\textbf{p}}_{view}, t}{\mathrm{argmin}}\, dist(h(\tilde{\textbf{p}}_{view}, \textbf{p}_{view}, \textbf{p}_{real}, t), \textbf{p}_{real}),\\
 s.t.&\,\,\,\, \tilde{\textbf{p}}_{view} \in \mathbb{P}_{seethru},\\
 h(\tilde{\textbf{p}}_{view}&, \textbf{p}_{view}, \textbf{p}_{real}, t) = \textbf{p}_{real}^+ + t \cdot (\textbf{p}_{real}^+ - \textbf{p}_{real}^-),\\
 \textbf{p}_{real}^+&, \textbf{p}_{real}^- = \tilde{f}_{seethru}(\tilde{\textbf{p}}_{view}, \frac{\textbf{p}_{view} - \tilde{\textbf{p}}_{view}}{\|\textbf{p}_{view} - \tilde{\textbf{p}}_{view}\|}),
  \label{eq:seethru_lifting_optimization}
\end{split}
\end{equation}
whose global optimal candidate $\tilde{\textbf{p}}_{view}^*$ is the projection of $\textbf{p}_{view}$. Thus $\textbf{p}_{view}$ will perceive $\textbf{p}_{real}$ in the same way that $\tilde{\textbf{p}}_{view}^*$ perceives it through the see-through optics. And the backward see-through model can be defined as
\begin{equation}
\begin{split}
  f^{-1}_{seethru} = \frac{\textbf{p}_{view} - \tilde{\textbf{p}}_{view}}{\|\textbf{p}_{view} - \tilde{\textbf{p}}_{view}\|},\\
  \tilde{\textbf{p}}_{view} = \pi_{seethru}^{\vee}(\textbf{p}_{view}, \textbf{p}_{real}),
  \label{eq:seethru_backward_constructed}
\end{split}
\end{equation}
which is simply the direction vector of the ray that passes through both $\textbf{p}_{view}$ and its projection on the base plane $\mathbb{P}_{seethru}$.

\section{Results}
\label{sec:results}

In this section, we will discuss how to incorporate our proposed models into the WLR system, including model calibration, integration with rendering frameworks, and evaluation of the models through simulation and testing on a real AR device.

\subsection{Model Calibration}
\label{sec:calibration}

In \cref{sec:main}, we introduced our proposed display and see-through models under the assumption that the intermediate models, defined in \cref{eq:display_intermediate_forward}, \cref{eq:display_intermediate_backward} and \cref{eq:seethru_forward_intermediate}, have been selected and fitted. In this section, we will provide more details on the intermediate model calibrations. Theoretically, any reasonable choice should be acceptable in our case and we choose the polynomial model for its simplicity and differentiability. Moreover, the calibration data (i.e. model training data) will be collected and used to minimize the model fitting error in order to determine the model parameters, which are polynomial coefficients in our setting. We will move on to the introduction of calibration data collection and skip the explanation of model fitting process, as polynomial fitting is a standard technique \cite{MLA:MKP2012}.

\begin{wrapfigure}{r}{0.2\textwidth}
    \centering
    \includegraphics[width=0.2\textwidth]{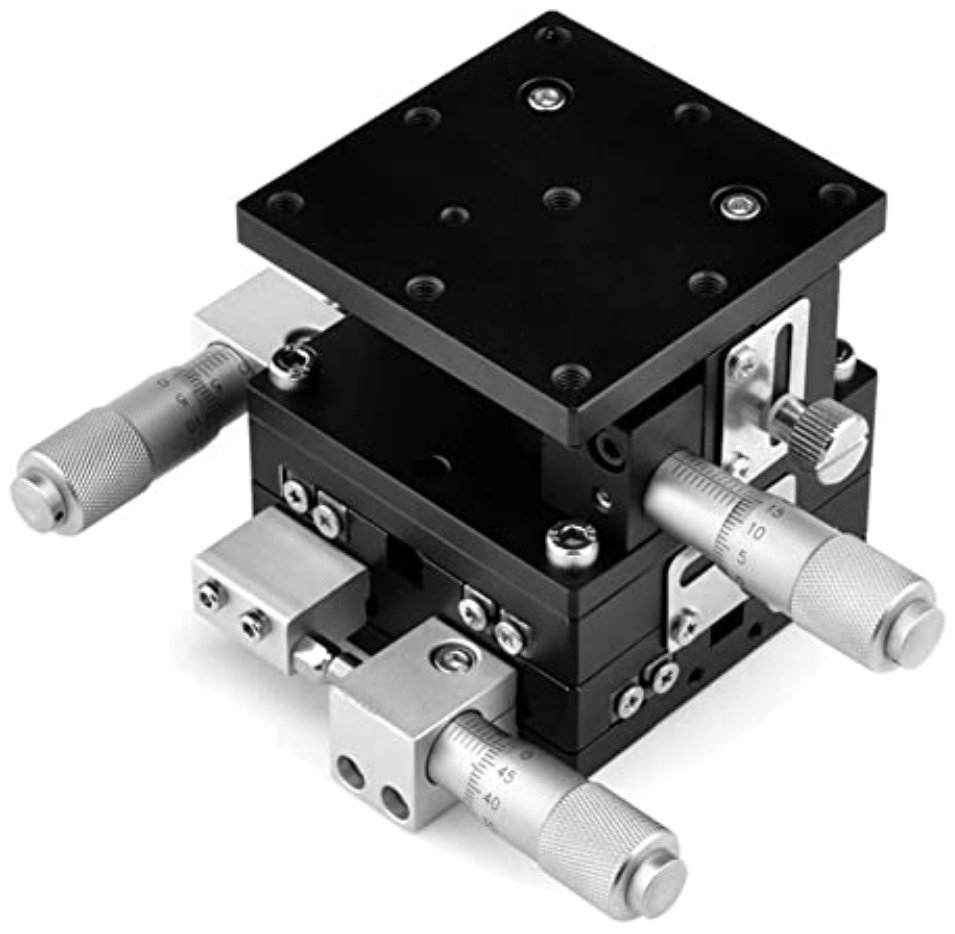}
    \caption{Motion stage with micrometer controller.}
    \label{fig:calibration_motion_stage}
\end{wrapfigure}

\paragraph{Display Model}
In \cref{eq:display_intermediate_forward} and \cref{eq:display_intermediate_backward}, we aim to model how a pixel $\textbf{p}_{pixel}$ is perceived from a viewpoint $\tilde{\textbf{p}}_{view}$ located on the base plane $\mathbb{P}_{display}$. As previously mentioned in \cref{sec:display}, we use a calibrated eyeball camera to capture the direction of a pixel's corresponding ray, $\textbf{v}_{view}$. The camera is rigidly mounted to a motion stage that is equipped with a micrometer controller as depicted in \cref{fig:calibration_motion_stage}. We build an HMD calibration station and position the camera around the eye box, which is located approximately 7-12 mm away from the display optics.

By adjusting the micrometer controller, we can move the camera to different locations within a single plane, which we refer to as the display base plane $\mathbb{P}_{display}$. At each camera position $\tilde{\textbf{p}}_{view}$, we collect correspondence data between display pixels $\textbf{p}_{pixel}$ and perceived rays $\textbf{v}_{view}$ by displaying some calibration patterns, such as gray code \cite{TGC:DRW2007}, fringe patterns \cite{PSA:ZFH2018}, etc. We can then fit the polynomial model for \cref{eq:display_intermediate_forward} and \cref{eq:display_intermediate_backward} using the collected data. To facilitate integration with the IOT subsystem, we convert the eyeball camera positions and perceived rays into the device frame before model fitting. We will wait until the next paragraph to explain how to perform the coordinate frame conversion.

\begin{figure*}[t]
     \centering
     \begin{subfigure}[b]{0.24\textwidth}
         \centering
         \includegraphics[width=\textwidth]{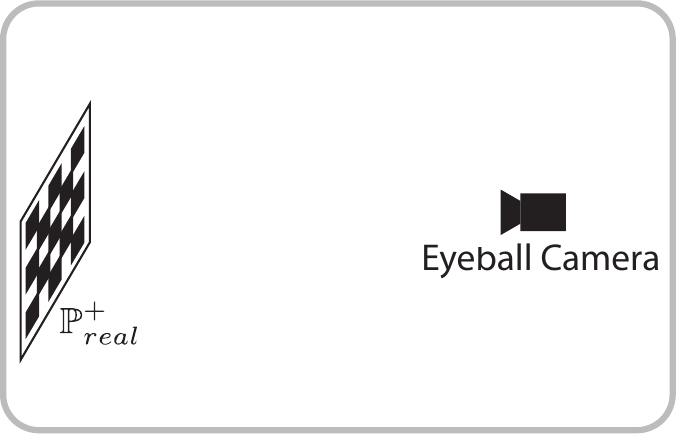}
         \caption{Two plane calibration.}
         \label{fig:two_plane_calibration_1}
     \end{subfigure}
     \hfill
     \begin{subfigure}[b]{0.24\textwidth}
         \centering
         \includegraphics[width=\textwidth]{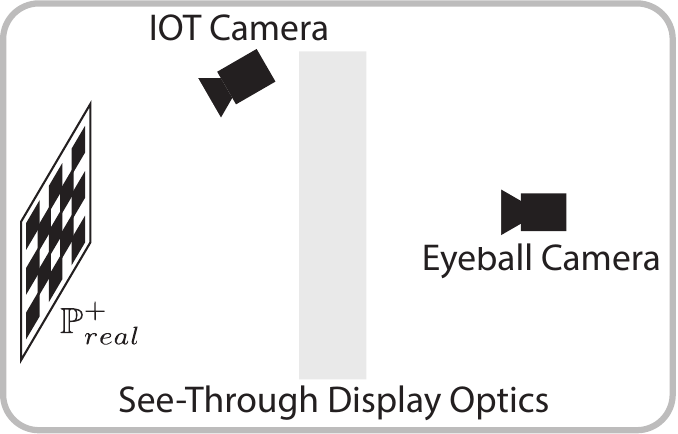}
         \caption{Two plane calibration.}
         \label{fig:two_plane_calibration_2}
     \end{subfigure}
     \hfill
     \begin{subfigure}[b]{0.24\textwidth}
         \centering
         \includegraphics[width=\textwidth]{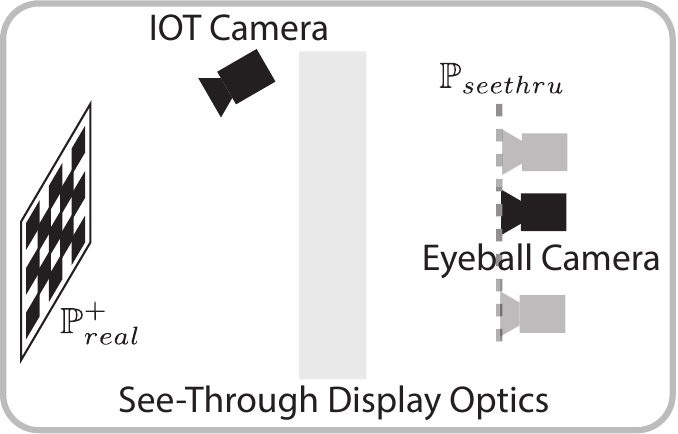}
         \caption{Two plane calibration.}
         \label{fig:two_plane_calibration_3}
     \end{subfigure}
     \hfill
     \begin{subfigure}[b]{0.254\textwidth}
         \centering
         \includegraphics[width=\textwidth]{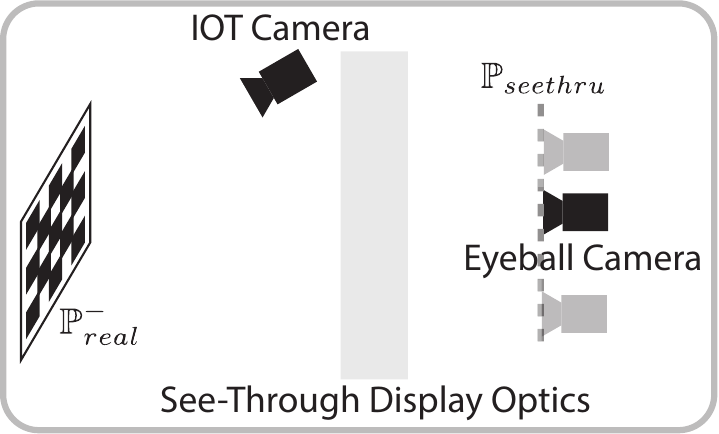}
         \caption{Two plane calibration.}
         \label{fig:two_plane_calibration_4}
     \end{subfigure}
        \caption{Two plane calibration.}
        \label{fig:two_plane_calibration}
\end{figure*}

\paragraph{See-Through Model}
\cref{eq:seethru_forward_intermediate} describes the correspondence between the light fields on the real-world side and the viewer side. We adopt a two-plane parameterization \cite{TLU:GGS1996,LFR:MP1996} for the real world light field and a base-direction parameterization for the other one. To collect the light field correspondence data, we follow these steps:
\begin{enumerate}
    \item Place a calibration target in front of the HMD station and use the eyeball camera to take one capture without the HMD mounted (see \cref{fig:two_plane_calibration}), which gives us the relative pose between the eyeball camera and the target plane $\mathbb{P}^+_{real}$.
    \item Put the HMD on the calibration station and use the IOT camera to take one capture of the calibration target (see \cref{fig:two_plane_calibration}), which gives us the pose of the eyeball camera and the target plane $\mathbb{P}^+_{real}$, both represented in the device frame.
    \item Use the micrometer controller to move the eyeball camera to a set of capture locations $\tilde{\textbf{p}}_{view}$ within the see-through base plane $\mathbb{P}_{seethru}$ (see \cref{fig:two_plane_calibration}). By using a pre-calibrated eyeball camera and matching the captured images with the calibration target pattern, we are able to obtain a collection of correspondences between the camera viewing directions $\textbf{v}_{view}$ and locations $\textbf{p}_{real}^+$ on the target plane for each camera position $\tilde{\textbf{p}}_{view}$, all represented in the device frame.
    \item Move the calibration target to a different location and repeat step 2 and 3 (see \cref{fig:two_plane_calibration}). From step 2, we have another target plane $\mathbb{P}_{real}^-$ represented in the device frame. From step 3, we collect the correspondences between $\textbf{v}_{view}$ and $\textbf{p}_{real}^-$ for the same set of $\tilde{\textbf{p}}_{view}$ (all in device frame) by repeating the camera capture locations.
\end{enumerate}

With the assumption that the micrometer controller is precise and the HMD calibration station remains stable when the HMD is mounted, we can collect the necessary fitting data for \cref{eq:seethru_forward_intermediate} using the steps described above.

\subsection{Rendering Framework Integration}
\label{sec:rendering}

Our calibrated display and see-through models can be integrated with the WLR system described in \cref{sec:background}. Now we outline the process of incorporating these models into two popular rendering frameworks, ray tracing and rasterization, used by such a system.

\paragraph{Ray Tracing}
In this framework, each pixel's value will be determined by tracing the path of its corresponding ray through a 3D scene \cite{PBR:PJH2016}. Given a viewpoint $\textbf{p}_{view}$ from the eye tracking system, our forward display and see-through models (\cref{eq:display_forward} and \cref{eq:seethru_forward}) can provide the necessary correspondence between display pixels $\textbf{p}_{pixel}$ and rays projected into the real world $(\textbf{p}^+_{real}, \textbf{p}^-_{real})$. Therefore, it is straightforward to use our models in combination with a ray tracing rendering engine to generate the display image.

\paragraph{Rasterization} This rendering technique works in a way that is different from ray tracing by projecting a 3D scene onto the image plane \cite{CGO:HBB2004}. We use a two-phase approach to integrate it with our models. The first step involves rendering an intermediate image by positioning the rasterization camera at a given viewpoint. It is then warped onto the display image space to generate the final result. To compute the mapping between display image space and intermediate image space, we use our forward display and see-through models to find the corresponding ray, projected into the real world, for each display pixel and intersect it with the 3D scene. The intersection is then mapped onto the intermediate image space, using the rasterization camera's projection, to obtain the corresponding 2D position for each display pixel. It is notable that using rasterization rendering engines will introduce errors to the WLR system because their camera projections do not take into account changes in the ray path caused by the see-through optics.

\subsection{Simulation Experiments}
\label{sec:simulation}

We perform Monte Carlo simulations to assess the individual performance of the display and see-through models. The CAD descriptions of a see-through display optical system serve as input for generating ground truth data through ray tracing. To obtain information for fitting and testing the models, we use a pinhole camera positioned at multiple sample locations. We sample the camera's image space to gather rays corresponding to camera pixels, and trace these rays through the optical system to find the matching display pixels for the display problem or the rays projected into the real world for the see-through problem. To mimic the system noise that occurs during actual data capture, we introduce zero-mean Gaussian perturbations to the camera pixel samples \cite{RCF:GTS2020}. The model accuracy is measured by quantifying the pixel offset for the backward display problem and the viewing angle discrepancy for the others \cite{SDA:IK2015}.

\subsection{On-Device Experiments}
\label{sec:on_device}

\begin{figure*}
     \centering
     \begin{subfigure}[b]{\textwidth}
         \centering
         \includegraphics[width=\textwidth]{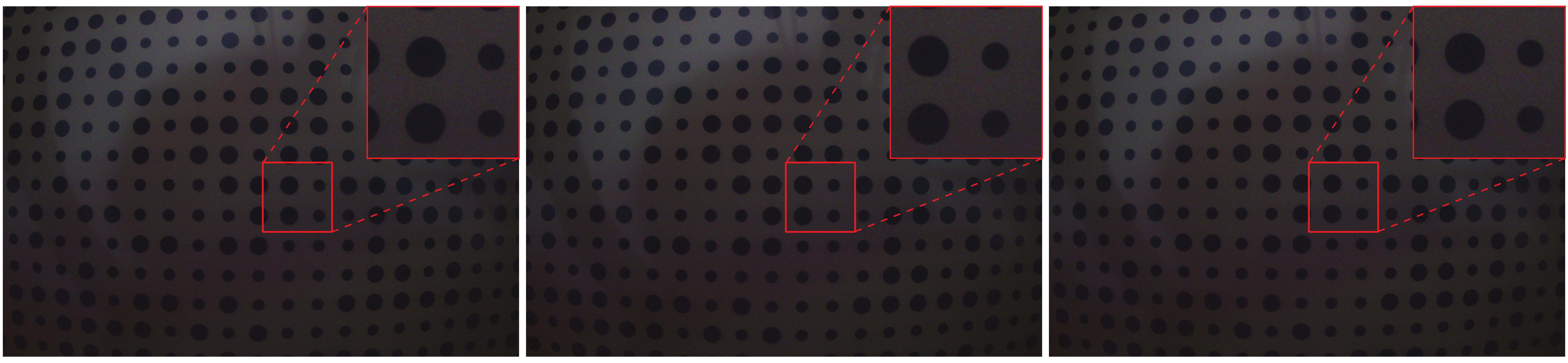}
         \caption{Calibu render experiment.}
         \label{fig:calibu_render_1}
     \end{subfigure}
     \hfill
     \begin{subfigure}[b]{\textwidth}
         \centering
         \includegraphics[width=\textwidth]{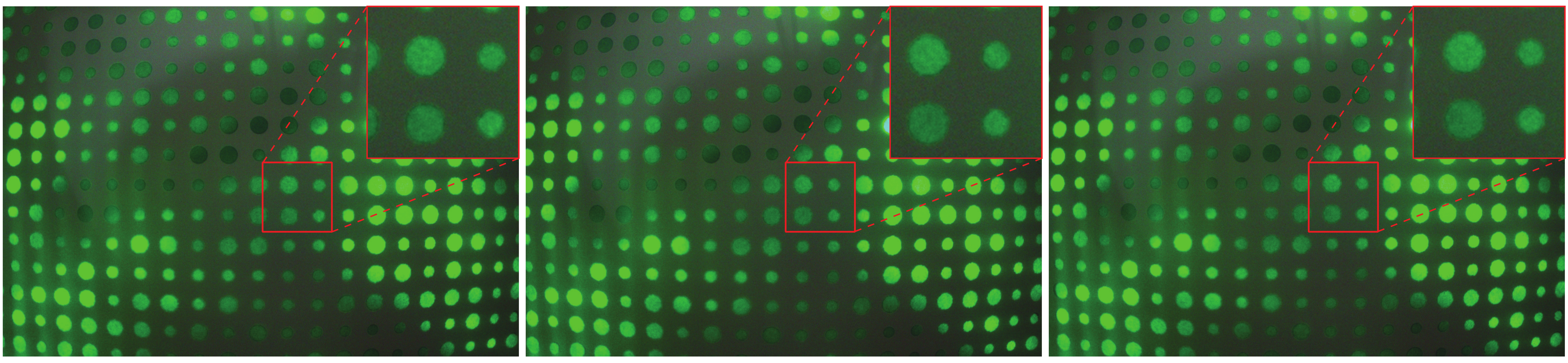}
         \caption{Calibu render experiment.}
         \label{fig:calibu_render_2}
     \end{subfigure}
        \caption{Calibu render experiment.}
        \label{fig:calibu_render}
\end{figure*}

We employ an HMD with see-through display optics and IOT system to evaluate the performance of display and see-through models. As described in \cref{sec:calibration}, we use a pair of pre-calibrated eyeball cameras, mounted on a motion stage with micrometer controller, to capture display and see-through calibration patterns. During the calibration phase for the display model, we will secure the HMD onto the calibration station and use the eyeball cameras to record the displayed gray code patterns \cite{TGC:DRW2007} at distinct locations within the 3D eye box region for volumetric sampling approach and different positions on a 2D plane in front of the display optics for our planar sampling method. In order to calibrate the see-through models, we use the same sampling positions for the eyeball cameras, which were utilized during the display calibration procedure. We use Calibu patterns \cite{RCF:GTS2020}, which are composed of a matrix of large and small dots arranged on a flat surface, to calibrate the see-through models. These dots, differentiated by the size of their adjacent dots, facilitate the efficient establishment of correspondences. 


To evaluate the precision of the calibrated models, we carry out an end-to-end verification procedure by integrating them with a ray tracing rendering engine and our IOT system. In a manner similar to the see-through calibration setup, we will position a Calibu pattern in front of the HMD calibration station and adjust the position of the eyeball cameras while the HMD is mounted. The IOT system will first calculate the pose of the Calibu pattern and transforms it into the device frame. Then the display and see-through models will work together to determine the image to be displayed as discussed in \cref{sec:rendering}, ensuring that the rendered Calibu pattern precisely aligns with the actual pattern from the perspective of each eyeball camera's position (see \cref{fig:calibu_render}). In our experiment, we position the Calibu pattern in distinct locations in front of the calibration station and capture images using the eyeball cameras from different viewpoints within the eye box region. For each static configuration of the Calibu pattern and eyeball cameras, we will take two pictures using the eyeball cameras. As depicted in \cref{fig:calibu_render}, one image will be captured with the display turned off, while the other will be captured with the display turned on. 
In this way, we can easily match the corresponding dots in each image pair and calculate the misalignment to evaluate the end-to-end model performance. 


{\small
\bibliographystyle{ieee_fullname}
\bibliography{egbib}

\begin{thebibliography}{10}\itemsep=-1pt

\bibitem{ISA:AB1994}
Ronald Azuma and Gary Bishop.
\newblock Improving static and dynamic registration in an optical see-through
  hmd.
\newblock In {\em Proceedings of the 21st Annual Conference on Computer
  Graphics and Interactive Techniques}, pages 197--204, 1994.

\bibitem{ASO:ART1997}
Ronald~T Azuma.
\newblock A survey of augmented reality.
\newblock {\em Presence: Teleoperators \& Virtual Environments}, 6(4):355--385,
  1997.

\bibitem{ODO:CKA2002}
Joshua~M Cobb, David Kessler, and John~A Agostinelli.
\newblock Optical design of a monocentric autostereoscopic immersive display.
\newblock In {\em International Optical Design Conference 2002}, volume 4832,
  pages 80--90. SPIE, 2002.

\bibitem{TGC:DRW2007}
Robert~W Doran.
\newblock The gray code.
\newblock Technical report, Department of Computer Science, The University of
  Auckland, New Zealand, 2007.

\bibitem{TCO:F1962}
Glenn~A Fry.
\newblock The center of rotation of the eye.
\newblock {\em American Journal of Optometry}, 39:581--595, 1962.

\bibitem{FCF:FSP1999}
Anton Fuhrmann, Dieter Schmalstieg, and Werner Purgathofer.
\newblock Fast calibration for augmented reality.
\newblock In {\em Proceedings of the ACM Symposium on Virtual Reality Software
  and Technology}, pages 166--167, 1999.

\bibitem{OST:GSW2000}
Yakup Genc, Frank Sauer, Fabian Wenzel, Mihran Tuceryan, and Nassir Navab.
\newblock Optical see-through hmd calibration: A stereo method validated with a
  video see-through system.
\newblock In {\em Proceedings IEEE and ACM International Symposium on Augmented
  Reality}, pages 165--174. IEEE, 2000.

\bibitem{PSF:GTN2002}
Yakup Genc, Mihran Tuceryan, and Nassir Navab.
\newblock Practical solutions for calibration of optical see-through devices.
\newblock In {\em Proceedings. International Symposium on Mixed and Augmented
  Reality}, pages 169--175, 2002.

\bibitem{VOF:GGW2018}
Ying Geng, Jacques Gollier, Brian Wheelwright, Fenglin Peng, Yusufu Sulai,
  Brant Lewis, Ning Chan, Wai Sze~Tiffany Lam, Alexander Fix, Douglas Lanman,
  and Others.
\newblock Viewing optics for immersive near-eye displays: Pupil swim/size and
  weight/stray light.
\newblock In {\em Digital Optics for Immersive Displays}, volume 10676, pages
  19--35. SPIE, 2018.

\bibitem{SCO:GFG2008}
Stuart~J Gilson, Andrew~W Fitzgibbon, and Andrew Glennerster.
\newblock Spatial calibration of an optical see-through head-mounted display.
\newblock {\em Journal of Neuroscience Methods}, 173(1):140--146, 2008.

\bibitem{TLU:GGS1996}
Steven~J Gortler, Radek Grzeszczuk, Richard Szeliski, and Michael~F Cohen.
\newblock The lumigraph.
\newblock In {\em Proceedings of The 23rd Annual Conference on Computer
  Graphics and Interactive Techniques}, pages 43--54, 1996.

\bibitem{ASO:GIM2018}
Jens Grubert, Yuta Itoh, Kenneth Moser, and J.~Edward Swan.
\newblock A survey of calibration methods for optical see-through head-mounted
  displays.
\newblock {\em IEEE Transactions on Visualization and Computer Graphics},
  24(9):2649--2662, 2018.

\bibitem{CUS:GTM2010}
Jens Grubert, Johannes Tuemle, Ruediger Mecke, and Michael Schenk.
\newblock Comparative user study of two see-through calibration methods.
\newblock {\em IEEE Virtual Reality (VR)}, 10(269-270):16, 2010.

\bibitem{PRF:GMS2022}
Phillip Guan, Olivier Mercier, Michael Shvartsman, and Douglas Lanman.
\newblock Perceptual requirements for eye-tracked distortion correction in vr.
\newblock In {\em ACM SIGGRAPH 2022 Conference Proceedings}, pages 1--8, 2022.

\bibitem{RCF:GTS2020}
Qi Guo, Huixuan Tang, Aaron Schmitz, Wenqi Zhang, Yang Lou, Alexander Fix,
  Steven Lovegrove, and Hauke~Malte Strasdat.
\newblock Raycast calibration for augmented reality hmds with off-axis
  reflective combiners.
\newblock In {\em 2020 IEEE International Conference on Computational
  Photography (ICCP)}, pages 1--12. IEEE, 2020.

\bibitem{MVG:HZ2003}
Richard Hartley and Andrew Zisserman.
\newblock {\em Multiple View Geometry in Computer Vision}.
\newblock Cambridge University Press, 2003.

\bibitem{CGO:HBB2004}
Donald Hearn, M~Pauline Baker, and M~Pauline Baker.
\newblock {\em Computer Graphics with OpenGL}, volume~3.
\newblock Pearson Prentice Hall Upper Saddle River, NJ:, 2004.

\bibitem{NDF:HSI2022}
Yuichi Hiroi, Kiyosato Someya, and Yuta Itoh.
\newblock Neural distortion fields for spatial calibration of wide
  field-of-view near-eye displays.
\newblock {\em Optics Express}, 30(22):40628--40644, 2022.

\bibitem{ETA:HNA2011}
Kenneth Holmqvist, Marcus Nystr{\"o}m, Richard Andersson, Richard Dewhurst,
  Halszka Jarodzka, and Joost Van~de Weijer.
\newblock {\em Eye Tracking: A Comprehensive Guide to Methods and Measures}.
\newblock OUP Oxford, 2011.

\bibitem{FB:HD1996}
Dale Husemoller.
\newblock {\em Fibre bundles}, volume~5.
\newblock Springer, 1966.

\bibitem{IFC:IK2014}
Yuta Itoh and Gudrun Klinker.
\newblock Interaction-free calibration for optical see-through head-mounted
  displays based on 3d eye localization.
\newblock In {\em 2014 IEEE Symposium on 3d User Interfaces (3DUI)}, pages
  75--82. IEEE, 2014.

\bibitem{PAS:IK2014}
Yuta Itoh and Gudrun Klinker.
\newblock Performance and sensitivity analysis of indica: Interaction-free
  display calibration for optical see-through head-mounted displays.
\newblock In {\em 2014 IEEE International Symposium on Mixed and Augmented
  Reality (ISMAR)}, pages 171--176. IEEE, 2014.

\bibitem{LFC:IK2015}
Yuta Itoh and Gudrun Klinker.
\newblock Light-field correction for spatial calibration of optical see-through
  head-mounted displays.
\newblock {\em IEEE Transactions on Visualization and Computer Graphics},
  21(4):471--480, 2015.

\bibitem{SDA:IK2015}
Yuta Itoh and Gudrun Klinker.
\newblock Simultaneous direct and augmented view distortion calibration of
  optical see-through head-mounted displays.
\newblock In {\em 2015 IEEE International Symposium on Mixed and Augmented
  Reality}, pages 43--48. IEEE, 2015.

\bibitem{TIA:ILS2021}
Yuta Itoh, Tobias Langlotz, Jonathan Sutton, and Alexander Plopski.
\newblock Towards indistinguishable augmented reality: A survey on optical
  see-through head-mounted displays.
\newblock {\em ACM Computing Surveys (CSUR)}, 54(6):1--36, 2021.

\bibitem{GCO:KBB2012}
Falko Kellner, Benjamin Bolte, Gerd Bruder, Ulrich Rautenberg, Frank Steinicke,
  Markus Lappe, and Reinhard Koch.
\newblock Geometric calibration of head-mounted displays and its effects on
  distance estimation.
\newblock {\em IEEE Transactions on Visualization and Computer Graphics},
  18(4):589--596, 2012.

\bibitem{RTI:KBB2018}
Kangsoo Kim, Mark Billinghurst, Gerd Bruder, Henry Been-Lirn Duh, and Gregory~F
  Welch.
\newblock Revisiting trends in augmented reality research: A review of the 2nd
  decade of ismar (2008--2017).
\newblock {\em IEEE Transactions on Visualization and Computer Graphics},
  24(11):2947--2962, 2018.

\bibitem{NPC:KSH2016}
Martin Klemm, Fabian Seebacher, and Harald Hoppe.
\newblock Non-parametric camera-based calibration of optical see-through
  glasses for ar applications.
\newblock In {\em 2016 International Conference on Cyberworlds (CW)}, pages
  33--40. IEEE, 2016.

\bibitem{HAP:KSH2017}
Martin Klemm, Fabian Seebacher, and Harald Hoppe.
\newblock High accuracy pixel-wise spatial calibration of optical see-through
  glasses.
\newblock {\em Computers \& Graphics}, 64:51--61, 2017.

\bibitem{LFR:MP1996}
Marc Levoy and Pat Hanrahan.
\newblock Light field rendering.
\newblock In {\em Proceedings of The 23rd Annual Conference on Computer
  Graphics and Interactive Techniques}, pages 31--42, 1996.

\bibitem{HPC:LM2013}
Mingyang Li and Anastasios~I Mourikis.
\newblock High-precision, consistent ekf-based visual-inertial odometry.
\newblock {\em The International Journal of Robotics Research}, 32(6):690--711,
  2013.

\bibitem{AMF:MT1999}
Erin McGarrity and Mihran Tuceryan.
\newblock A method for calibrating see-through head-mounted displays for ar.
\newblock In {\em Proceedings 2nd IEEE and ACM International Workshop on
  Augmented Reality}, pages 75--84. IEEE, 1999.

\bibitem{AMS:MRO2007}
Anastasios~I Mourikis, Stergios~I Roumeliotis, et~al.
\newblock A multi-state constraint kalman filter for vision-aided inertial
  navigation.
\newblock In {\em ICRA}, volume~2, page~6, 2007.

\bibitem{MLA:MKP2012}
Kevin~P Murphy.
\newblock {\em Machine Learning: A Probabilistic Perspective}.
\newblock MIT press, 2012.

\bibitem{DRC:OZA2004}
Charles~B Owen, Ji Zhou, Arthur Tang, and Fan Xiao.
\newblock Display-relative calibration for optical see-through head-mounted
  displays.
\newblock In {\em Third IEEE and ACM International Symposium on Mixed and
  Augmented Reality}, pages 70--78. IEEE, 2004.

\bibitem{PBR:PJH2016}
Matt Pharr, Wenzel Jakob, and Greg Humphreys.
\newblock {\em Physically Based Rendering: From Theory to Implementation}.
\newblock Morgan Kaufmann, 2016.

\bibitem{OCM:SHY2019}
Kiyosato Someya, Yuichi Hiroi, Makoto Yamada, and Yuta Itoh.
\newblock Ostnet: Calibration method for optical see-through head-mounted
  displays via non-parametric distortion map generation.
\newblock In {\em 2019 IEEE International Symposium on Mixed and Augmented
  Reality Adjunct (ISMAR-Adjunct)}, pages 259--260. IEEE, 2019.

\bibitem{PR:TS2002}
Sebastian Thrun.
\newblock Probabilistic robotics.
\newblock {\em Communications of the ACM}, 45(3):52--57, 2002.

\bibitem{SPA:TN2000}
Mihran Tuceryan and Nassir Navab.
\newblock Single-point active alignment method (spaam) for optical see-through
  hmd calibration for augmented reality.
\newblock In {\em Proceedings IEEE and ACM International Symposium on Augmented
  Reality}, volume~11, pages 149--158, 2000.

\bibitem{PSA:ZFH2018}
Chao Zuo, Shijie Feng, Lei Huang, Tianyang Tao, Wei Yin, and Qian Chen.
\newblock Phase shifting algorithms for fringe projection profilometry: A
  review.
\newblock {\em Optics and Lasers in Engineering}, 109:23--59, 2018.

\end{thebibliography}
}

\end{document}